\documentclass[12pt]{article}
\usepackage{sc3conf}
\usepackage{amsfonts}
\usepackage{epsfig}

\newcommand{\rhof}{\rho_{\rm f}}
\newcommand{\rhos}{\rho_{\rm s}}

\newcommand{\gcc}{g cm$^{-3}$ }

\def\la{\;\raise0.3ex\hbox{$<$\kern-0.75em\raise-1.1ex\hbox{$\sim$}}\;}
\def\ga{\;\raise0.3ex\hbox{$>$\kern-0.75em\raise-1.1ex\hbox{$\sim$}}\;}
\def\lr{\;\raise0.3ex\hbox{$\rightarrow$\kern-1.0em\raise-1.1ex\hbox{$\leftarrow$}}\;}

\begin{document}
\raggedbottom

\title{Pycnonuclear burning and accreting neutron stars}

\authors{D.G.\ Yakovlev 
    and K.P.\ Levenfish 
}
\addresses{
   Ioffe Physical Technical Institute, 194021 St.-Petersburg, Russia\\
   e-mails: yak@astro.ioffe.rssi.ru, ksen@astro.ioffe.rssi.ru
}

\maketitle

\begin{abstract}
We outline the phenomenon of deep crustal heating 
in transiently accreting
neutron stars. It is produced by nuclear transformations
(mostly, by pycnonuclear reactions)
in accreted matter while this matter sinks to
densities $\rho \ga 10^{10}$
g cm$^{-3}$ under the weight of freshly accreted material.
We consider then
thermal states of transiently accreting neutron
stars (with mean mass accretion rates $\dot{M} \sim 10^{-14}-10^{-9}$
M$_\odot$/yr) determined by deep crustal heating.
In a simplified fashion we study how the thermal flux
emergent from such stars depends on the properties
of superdense matter in stellar interiors.
We analyze the most important regulators of the thermal flux:
strong superfluidity in the cores of low-mass stars
and fast neutrino emission
(in nucleon, pion-condensed, kaon-condensed, 
or quark phases of dense matter)
in the cores of high-mass stars. We compare the results
with observations of soft X-ray transients in quiescent states.
\end{abstract}


\section{Introduction}
\label{introduction}

Nuclear burning of matter is the keystone of stellar physics.
It is vitally important
for main-sequence stars where it proceeds in {\it thermonuclear}
regime. In this regime, Coulomb barrier in nuclear reactions
is penetrated owing to the thermal energy of colliding nuclei.

In the present paper, we discuss another,
{\it pycnonuclear} regime of nuclear
burning. Coulomb barrier in pycnonuclear
reactions is penetrated due to (quantum) zero-point motion of
the reacting nuclei. While the thermonuclear regime
is realized in sufficiently low-density and warm plasma
(centers of main-sequence stars as an example),
the pycnonuclear regime operates at high densities and
not too high temperatures. The pycnonuclear reaction rates
are almost temperature independent, i.e.,
this burning occurs even at $T=0$. The formalism of
nuclear reactions in
the pycnonuclear regime, as well as in the intermediate
thermo-pycnonuclear
regimes, was developed in a
seminal paper by Salpeter \& van Horn \cite{svh69}.
The pycnonuclear reactions
have been considered later in a number of publications
(e.g., \cite{sc90,kitamura00} and references therein).

We analyse implication of the concept of pycnonuclear
burning to the physics of neutron stars (NSs). They are
compact stellar objects of mass $M \sim 1.4 \, {\rm M}_\odot$
(M$_\odot$ being the solar mass)
and radius $R \sim 10$ km. The density of matter in their cores
reaches $\sim 10^{15}$ g cm$^{-3}$ (i.e., several times
nuclear matter density). The composition
and equation of state of the NS cores cannot be
unambiguously calculated and is still
almost not constrained by observations (see, e.g., Ref.\ \cite{lp01}).
For instance, inner NS cores may contain
nucleon matter, pion- or kaon condensates, or quark matter.
Neither of this hypothesis can be accepted
or rejected at present. Neutrons, protons, and
other strongly interacting species can be in
superfluid state. Microscopic calculations predict
density dependent critical temperatures of
these superfluids in the range $10^8-10^{10}$ K or
higher (e.g., \cite{yls99,ls01} and references therein). 
Such calculations are very model dependent
and introduce additional uncertainty into the 
physics of NS interiors.

The internal structure of NSs is studied
by confronting theoretical models with observations
in many ways, for instance, basing on precisely measured masses
of some radio pulsars in binary systems (e.g., \cite{tc99,lp01}),
or comparing the theory and observations of cooling NSs
(e.g., \cite{page98,ygkp02,yh02}). 
Pycnonuclear
burning of matter in transiently accreting NSs
gives another method to study the internal structure
of NSs as described below.

\section{Pycnonuclear burning and deep crustal heating}

Let us follow the evolution of accreted matter in
a transiently accreting NS (in a binary system).
An infall of accreted matter is accompanied by a
large energy release ($\sim 300$ MeV per accreting
nucleon) due to the transformation of infall energy
into heat. This heat is most likely radiated away
by photons from the NS surface and cannot
warm up the star. An accreted material sinks then gradually into
the NS interior under the weight of newly infalling
matter. In the outer stellar layer (at densities
$\rho \la 10^{10}$ g cm$^{-3}$, a few tens of meters under
the surface) this matter burns (mainly in the thermonuclear
regime, in a steady or explosive manner) into heavier
elements, to Fe. The energy released in this burning
is mostly transferred by thermal conductivity
to the surface and again radiated away by photons,
producing no heating of NS interiors.

Let us focus on subsequent transformations of
accreted matter at $10^{10} \la \rho$ $ \la 10^{13}$ g cm$^{-3}$.
They include: beta captures, absorption and
emission of neutrons, and pycnonuclear reactions.
These transformations and associated energy release
have been considered in detail by Haensel \& Zdunik \cite{hz90}
(see the monograph \cite{bisnovatyi} for
references to some earlier work).
The appropriate reaction rates do not depend on temperature
but strongly depend on $\rho$. It is sufficient
to assume that any transformation occurs in a certain
infinitesimally thin NS layer.
The main energy release (0.79 MeV per accreting baryon)
takes place \cite{hz90} at densities from about $10^{12}$
to $10^{13}$ g cm$^{-3}$, about 1 km under the
surface, in three pycnonuclear reactions.
The total energy release is about 1.45 MeV per
accreting baryon.
The total heating power is determined by the mass accretion rate
$\dot{M}$ and estimated as
\begin{equation}
  L_{\rm dh} = 1.45~{\rm MeV} \, \dot{M}/m_{\rm N}
  \approx 8.74 \times 10^{33}  \, 
  \dot{M}/(10^{-10} \, {\rm M}_\odot \; {\rm yr}^{-1})
  \;\;{\rm   erg\; s}^{-1},
\label{Ldh} 
\end{equation}
where 
$m_{\rm N}$ is the nucleon mass.
This energy release produces {\it deep crustal heating}
of accreted matter. In contrast to (much stronger) heating
near the surface,
this heat is spread by thermal conductivity 
over the entire NS and warms it up.

\section{Soft X-ray transients}

It is likely that deep crustal heating manifests itself
in soft X-ray transients (SXRTs). We mean
SXRTs containing NSs
in binary systems with low-mass companions
(low-mass X-ray binaries) \cite{csl97}.
Such objects undergo the periods of outburst activity
(days--months, sometimes years)
superimposed with
the periods of quiescence (months--decades). This transient activity is
most probably regulated
by accretion from disks surrounding the NSs.
An active period is associated with a switched-on
accretion. The accretion energy released at the
NS surface is large for a transient
to look like a bright X-ray source
with the luminosity $L_X \sim 10^{36}-10^{38}$
erg s$^{-1}$. The accretion is
switched off or strongly suppressed during quiescence periods when
$L_X$ drops to $L_X \la 10^{34}$
erg~s$^{-1}$. 

The nature of the quiescent emission is still uncertain.
The hypothesis that this emission
is produced by the thermal flux emergent from the NS
interior has been rejected initially due to
two reasons. First, the radiation spectra
fitted with the blackbody model have given unreasonably
small NS radii. Second,
the NSs in SXRTs have been expected
to be old and thus internally cold; their quiescent
emission should have been much lower than the observed one. 
These arguments were questioned by Brown et al.\ \cite{bbr98}
who suggested that the NSs can be warmed up 
by the deep crustal heating
while the radiation spectra can be fitted with the hydrogen
atmosphere models yielding realistic NS radii.
Since
the emergent radiation flux may depend on the
NS internal structure
this opens an attractive possibility 
(see \cite{ur01,cgpp01,rbbpzu02,bbc02}
and references therein)
to explore
the internal structure by comparing
observations of SXRTs in quiescence
with theoretical models.

\section{Thermal states of accreting neutron stars}

Let us outline the theory of thermal states
of transiently accreting NSs. Following \cite{yh02}
we will make a general but not very accurate
analysis of the problem.
The NSs of study will have internal temperatures
$T \la 3 \times 10^8$ K. They are thermally
inertial objects with
thermal relaxation times 
$\sim10^4$ yr \cite{cgpp01}.
The quiescence intervals in SXRTs
are much shorter than these relaxation times.
We neglect short-term variability
in the stellar crust; it can be associated \cite{ur01,bbc02}
with variable residual accretion in quiescence, thermal
relaxation of transient 
deep crustal heating, etc. 
Instead, we focus on a (quasi)stationary
steady state of the NS
determined by the accretion rate $\dot{M}\equiv \langle \dot{M} \rangle$ 
(from $10^{-14}$ to $10^{-9}$ M$_\odot$/yr)
averaged over time intervals comparable with the thermal
relaxation time.
Thus we replace variable $\dot{M}$ with $\langle \dot{M} \rangle$.
The accretion rates of study are too low 
to noticeably increase NS mass, $M$,
during long periods of SXRT evolution.

Thermal states of accreting NSs
can be found by solving the equation of thermal
balance  \cite{gs80}  in the approximation of thermally relaxed,
isothermal stellar interior 
enclosed by a thin ($\rho \la 10^{10}$ g cm$^{-3}$)
heat-blanketing envelope:
\begin{equation}
   C(T_i) \, {{\rm d}T_i \over {\rm d}t}=L_{\rm dh}^\infty(\dot{M})
       -L_\nu^\infty(T_i)-L_\gamma^\infty(T_{\rm s}).
\label{therm-balance}
\end{equation}
Here, $T_{\rm s}$ is the effective surface temperature,
$T_i(t)$ is the internal temperature
(constant throughout the isothermal interior),
$C$ is the NS total heat capacity,
$L_\nu^\infty$ is the neutrino luminosity,
$L_\gamma^\infty$ the photon luminosity,
and $L_{\rm dh}^\infty$ is the deep-heating power.
The relation between $T_i$ and $T_{\rm s}$ 
is determined (e.g., \cite{pcy97}) 
from the solution of thermal conduction
problem in the heat-blanketing envelope.
Equation (\ref{therm-balance}) takes proper
account of the effects of General Relativity 
(see Ref.\ \cite{thorne77}, for details)
important in such compact objects as NSs.
Particularly, $T_{\rm s}$ refers to a local reference frame
on the NS surface while the effective temperature
detected by a distant observer is $T_{\rm s}^\infty=T_{\rm s} \,
\sqrt{1-r_{\rm g}/R}$, where $r_{\rm g}=2 G M/c^2$ is
the Schwarzschild radius, $M$ the gravitational NS mass,
$R$ the circumferential NS radius, and
$G$ is the gravitational constant. 
The internal temperature $T_i$ in Eq.\ (\ref{therm-balance}) also
refers to a distant observer, while the local internal temperature
is $T(r,t)=T_i(t)\,{\rm e}^{-\Phi(r)}$, $\Phi(r)$ being the
metric function, with ${\rm e}^{\Phi(R)}=\sqrt{1-r_{\rm g}/R}$.
The quantities $C$ and $L_\alpha^\infty$ ($\alpha=\nu$, $\gamma$,
dh) have to be calculated with account for General Relativity;
all three quantities $L_\alpha^\infty$ refer to a distant observer.
In particular, $L_\gamma^\infty=4 \pi R^2 \sigma T_{\rm s}^4
(1-r_{\rm g}/R)$, and 
$L_\nu^\infty=\int {\rm d}V \, Q_\nu \, {\rm e}^{2 \Phi}$,
where $Q_\nu$ is the neutrino emissivity and ${\rm d}V$ is
a proper volume element.
A steady-state accretion in General Relativity
is characterized by a constant mass accretion rate $\dot{M}$
which determines \cite{tz77} 
constant number of accreting baryons
passing through a sphere of any radial coordinate $r$
per unit time for a distant observer. 
For a deep crustal heating in a relatively thin
NS envelope one approximately has (e.g., Ref.\ \cite{ylh02})
$L_{\rm dh}^\infty = L_{\rm dh} \,
\sqrt{1-r_{\rm g}/R}$, where $L_{\rm dh}$ is given by
Eq.\ (\ref{Ldh}). Since $\dot{M}$ in SXRTs is determined
with large uncertainties, we
set $L_{\rm dh}^\infty = L_{\rm dh}$ in our subsequent
analysis.

As discussed above, we are interested in the steady-state
solution of Eq. (\ref{therm-balance}).
In this case it is sufficient to solve the 
simplified heat-balance equation
\begin{equation}
     L_{\rm dh}^\infty(\dot{M})=L_\nu^\infty(T_i) 
    + L_\gamma^\infty(T_{\rm s}),
\label{Main}
\end{equation}
where $L_{\rm dh}^\infty$ is known once $\dot{M}$
is specified. 
The solutions 
give us a {\it heating curve}, the dependence of
the photon luminosity
on the mean accretion rate, $L^\infty_\gamma(\dot{M})$.
It has been shown (e.g., Ref.\ \cite{ylh02})
that the theory of heating of accreting NSs is
very similar to theory of cooling of isolated NSs.

To solve Eq.\ (\ref{Main}) we employ
a simple toy model \cite{yh02} of the NS thermal structure.
It allows us to analyze various neutrino emission
scenarios in NS cores.
Let us remind the main points.

The toy model assumes that a NS core
is divided into three zones:
the {\it outer} zone, $\rho< \rho_{\rm s}$;
the {\it transition} zone, $\rho_{\rm s} \leq \rho < \rho_{\rm f}$;
and the {\it inner} zone, $\rho \geq \rho_{\rm f}$.
If the NS central density
$\rho_{\rm c} \leq \rho_{\rm s}$, two last zones are absent.

In the outer zone, the neutrino emission is supposed to be
{\it slow}, while in the inner zone it is
{\it fast}. The neutrino emissivity is assumed to be given by:
\begin{equation}
     Q_\nu^{\rm slow}(\rho \leq \rho_{\rm s})=Q_{\rm s} T_9^8,\qquad
     Q_\nu^{\rm fast}(\rho \geq \rho_{\rm f})=Q_{\rm f} T_9^6.
\label{Qs}
\end{equation}
Here, $T_9$ is the local internal stellar temperature $T$
in $10^9$ K, while $Q_{\rm s}$ and $Q_{\rm f}$
are constants.
For simplicity, the toy model uses the linear
interpolation in $\rho$ between
$Q_\nu^{\rm slow}$ and $Q_\nu^{\rm fast}$ in the transition zone.

This {\it generic} description of $Q_\nu$
covers many {\it physical} models
of nucleon and exotic supranuclear
matter with different leading neutrino processes
listed in Tables 1 and 2
(from Ref.\ \cite{yh02}). In these tables,
N is a nucleon (neutron or proton, n or p); e is an electron;
$\nu$ and $\bar{\nu}$ are neutrino and antineutrino;
q is a quasinucleon (mixed n and p states); u and d are quarks.

Slow neutrino processes are important in the outer
NS cores composed of nucleons and electrons where
the proton fraction is too low to allow for fast direct Urca (Durca)
process \cite{lpph91}.
In particular, $Q_{\rm s}$ can describe
modified Urca (Murca) process
in nonsuperfluid nucleon matter,
or weaker
NN-brems\-strah\-lung (e.g., nn-bremsstrahlung if Murca is
suppressed by a strong proton superfluidity as
considered in Ref.\ \cite{kyg02}).

Fast neutrino emission is produced in the inner NS cores.
Its intensity is regulated by the parameter $Q_{\rm f}$ and
depends
on the composition of dense matter. The fast emission may be produced 
either by
a very powerful Durca process in nucleon matter or
somewhat weaker similar
processes in exotic phases of
matter (pion condensed, kaon condensed,
or quark matter) \cite{pethick92,ykgh01}.
The bottom line of Table 2
refers to nonsuperfluid quark matter in NS cores.

\newcommand{\rrr}{\rule{0cm}{0.3cm}}

\begin{table}[t]
\caption{Main processes of slow neutrino emission
in nucleon matter: Murca and
bremsstrahlung (brems)}
\begin{center}
  \begin{tabular}{|lll|}
  \hline
  Process   &    &  $Q_{\rm s}$, erg cm$^{-3 \rrr}$ s$^{-1}$ \\
  \hline
  Murca &
  ${\rm nN \to pN e \bar{\nu} \quad
   pN e \to nN \nu } $ &
  $\quad 10^{20 \rrr}-3 \times 10^{21}$  \\
  Brems. &
  ${\rm NN \to NN  \nu \bar{\nu}}$  &
  $\quad 10^{19 \rrr}-10^{20}$\\
   \hline
\end{tabular}
\label{tab-nucore-slow}
\end{center}
\end{table}

\begin{table}[t]
\caption{Leading processes of fast
neutrino emission
in nucleon matter and three models of exotic
         matter}
\begin{center}
  \begin{tabular}{|lll|}
  \hline
  Model              & Process             &
        $Q_{\rm f}$, erg cm$^{-3 \rrr}$ s$^{-1}$ \\
  \hline
  Nucleon matter &
  ${\rm n \to p e \bar{\nu} \quad
   p e \to n \nu }$ & $\quad
  10^{26 \rrr}-10^{27}$  \\
  Pion condensate &
  ${\rm q \to q e \bar{\nu} \quad
   q e \to q {\nu} } $ & $ \quad
  10^{23 \rrr}-10^{26}$  \\
   Kaon condensate &
   ${\rm q \to q e \bar{\nu} \quad
    q e \to q \nu } $ & $ \quad
   10^{23 \rrr}-10^{24}$ \\
   Quark matter &
   ${\rm d \to u e \bar{\nu} \quad  u e \to d \nu } $ & $  \quad
   10^{23 \rrr}-10^{24}$ \\
   \hline
\end{tabular}
\label{tab-nucore-fast}
\end{center}
\end{table}

The toy model solves Eq.\ (\ref{Main})
under a number of simplified assumptions
\cite{yh02}.
The density profile in the star is approximate,
$\rho(r)=\rho_{\rm c}\, ( 1 - r^2/R^2)$, so that the NS mass
is $M=8 \pi R^3 \rho_{\rm c}/15$.
The effect of various
equations of state can be mimicked by choosing different
$M-R$ relations. For simplicity, following
Ref.\ \cite{yh02},
we set $R=12$ km and allow
the central density $\rho_{\rm c}$ to vary
from $7 \times 10^{14}$ to $1.4 \times 10^{15}$ g cm$^{-3}$,
varying thus $M$ from 1.02 M$_\odot$ to 2.04 M$_\odot$.
More realistic $M-R$ relations will not change our
principal conclusions. 

\section{Results and discussion}

The resulting heating curves are presented in Fig.\ 1.
For certainty, we consider two representative
NS models: a low-mass model
($\rho_{\rm c}=8\times 10^{14}$ g cm$^{-3}$,
$M=1.16\,{\rm M}_\odot$) and a high-mass model
($\rho_{\rm c}=1.4 \times 10^{15}$ g cm$^{-3}$, $M=2.04 \, {\rm M}_\odot$)
and adopt $\rhos = 8\times 10^{14}$ \gcc
and $\rhof=10^{15}$ \gcc. Thus the low-mass NS model
contains the outer core alone and is a typical
example of a NS with slow neutrino emission.
The high-mass model contains a bulky
inner core and is an example of a NS governed by fast neutrino
emission.

\begin{figure}[t]
\centering
\epsfxsize=13.8cm
\epsffile[55 180 495 555]{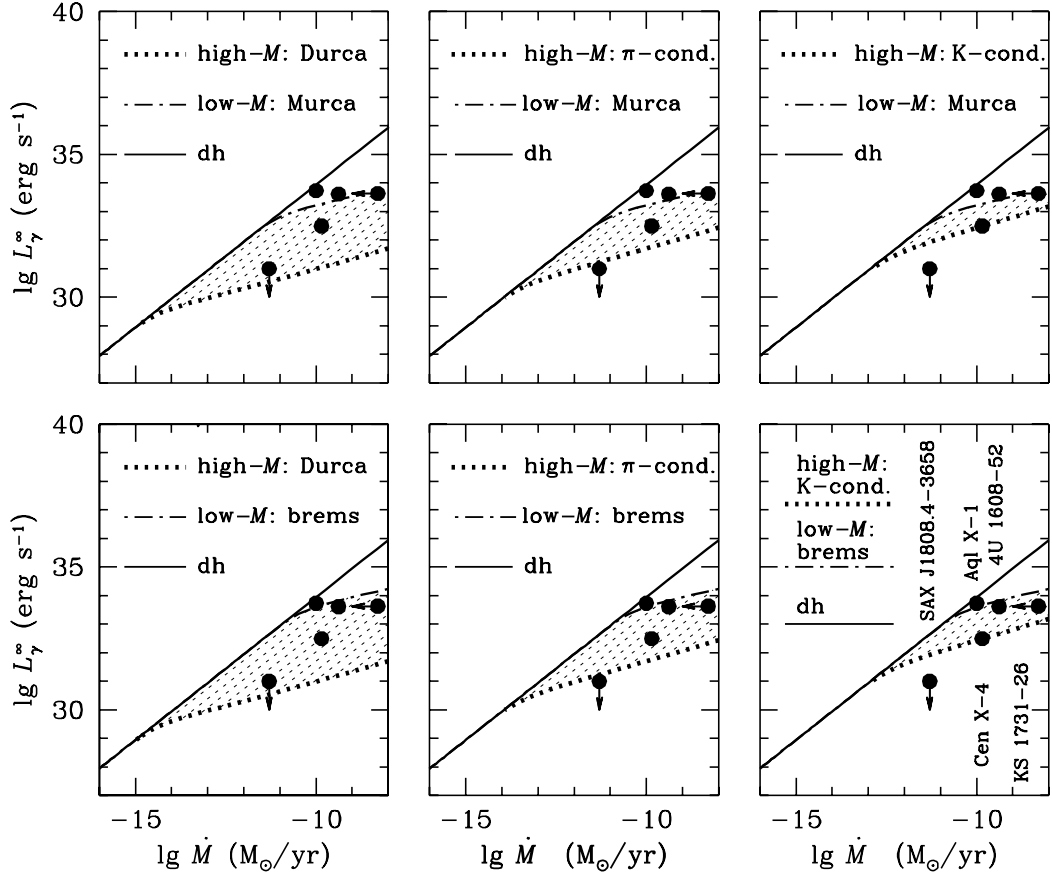}
\caption{\hspace{.2cm}
Allowable thermal states (dotted regions) of accreting NSs
for six physical pictures of NS interiors 
compared with observations of SXRTs.
}
\label{fig}
\end{figure}

Let us fix a {\it physical picture} of NS interiors
(a set of four parameters, $Q_{\rm s}$, $Q_{\rm f}$, $\rho_{\rm s}$,
and $\rho_{\rm f}$, in our case). It has to be the same
for all NSs, and we would like to constrain it from
observations of SXRTs. Figure 1 shows six representative
cases. 
Solid curves present the deep heating power, $L_{\rm dh}^\infty$,
which is the absolute upper limit of $L_\gamma^\infty$ (see Eq.\ (\ref{Main})).
The heating curves of low-mass NSs are shown by dash-and-dot
lines and the curves of high-mass NSs are shown by dotted lines. 
Three upper panels display
low-mass NSs with slow neutrino emission determined
by Murca process in nonsuperfluid NS cores ($Q_{\rm s}=10^{21}$).
Three lower panels display
low-mass stars with very slow neutrino emission
appropriate to neutron-neutron bremsstrahlung
in the NS cores with strong proton superfluidity 
($Q_{\rm s}= 3 \times 10^{19}$). 
The superfluidity 
damps Murca process
and enables us to obtain hotter
NSs, just as in the theory of cooling NSs
\cite{kyg02,ygkp02}.
Two left panels exhibit  
high-mass NSs with 
the fastest neutrino emission from 
nucleon NS cores with open Durca process ($Q_{\rm f}=10^{27}$).
Two middle panels show high-mass NSs with
pion-condensed inner cores ($Q_{\rm f}=10^{25}$),
while two right panels show high-mass NSs with
kaon-condensed or quark cores ($Q_{\rm f}=10^{23}$).

A heating curve of a low-mass NS (with given $Q_{\rm s}$)
provides an upper 
limit of $L_\gamma^\infty$, 
whereas a heating curve of a high-mass NS (with given $Q_{\rm f}$)
gives a lower limit of $L_\gamma^\infty$,
for a fixed physical picture of NS interiors.  
Varying the NS mass from the lower values to the higher
we obtain a family of heating curves which fill in the 
(dotted) space
between the upper and lower curves.
A group of NSs whose heating curves lie essentially
between the upper and lower curves will be called {\it medium-mass}
stars. Their mass range 
is sensitive to the position and
width of the transition layer ($\rho_{\rm s} \la \rho
\la \rho_{\rm f}$) between the slow and fast neutrino
emission zones, in full analogy with the theory
of cooling NSs \cite{kyg02,ygkp02,yh02}.

As seen from Fig.\ 1 and Eq.\ (\ref{Main}),
there are two limiting regimes. 

{\it Photon emission regime}\/ is realized in cold enough NSs, where 
$\dot{M}$ is rather low and $L_\nu^\infty \ll L_\gamma^\infty$.
It is equivalent to the photon emission stage of 
cooling NSs. In this case Eq.\ (\ref{Main}) reduces to:
$L^\infty_{\rm dh}(\dot{M}) = L_\gamma^\infty(T_{\rm s})$, 
i.e., all the heat released in the 
deep crust diffuses to the surface
and is emitted away by photons.
Accordingly, the photon luminosity is determined by the 
mass accretion rate and is independent of the NS internal structure.

{\it  Neutrino  emission regime}\/ is realized in warmer  NSs, where
$\dot{M}$ is sufficiently high and
$L_\nu^\infty \gg L_\gamma^\infty$, i.e., the 
deep crustal heat is spread by thermal diffusion over the star
and carried away by neutrino emission. 
This regime is equivalent to 
neutrino stage in cooling NSs. The thermal balance equation then reads:
$L^\infty_{\rm dh}(\dot{M}) = L_\nu^\infty(T_i)$, which gives $T_i$.
The surface photon luminosity just responds to $T_i$
and depends on the NS internal structure.

Analytic estimates of $T_{\rm s}$ in
both regimes are given in Ref.\ \cite{ylh02}. As seen
from Fig.\ 1, a transition from the photon to neutrino
regime with increasing $\dot{M}$ takes place at
$\dot{M} \sim 10^{-15}$, $10^{-14}$, and $10^{-13}$
M$_\odot$/yr, for high-mass NS models with nucleon Durca,
pion-condensed, and kaon-condensed inner cores, respectively.

Thus, for any physical picture of NS interiors we 
obtain its own upper and lower heating curves,
and intermediate heating curves of medium-mass NSs.
These results can be confronted with observations of SXRTs
containing NSs.  Figure 1
presents an example of such an analysis
for five SXRTs. Following Ref.\ \cite{ylh02} we take
the observational
data  for Aql X--1, Cen X--4, 4U 1608--522, 
KS 1731--26, and SAX J1808.4--3658 from Refs.\ 
\cite{rbbpz02,rbbpz00},
\cite{rbbpz01,csl97}, 
\cite{rbbpz99,csl97}, 
\cite{wgvm02,rbbpzu02}, 
and \cite{campanaetal02,bc01}, respectively.
We regard $L_\gamma^\infty$ as the thermal surface
luminosity of these sources in quiescent states,
and we regard $\dot{M}$ as the
mass accretion rate averaged \cite{ylh02} 
over representative intervals of time. These time intervals
include quiescent and active periods (while the accreted mass
is mainly accumulated in the active states).
The value of $\dot{M}$ for KS 1731--26
is most probably an upper limit. No quiescent   thermal
emission has been detected from SAX J1808.4--3658, 
and we present the upper limit of $L_\gamma^\infty$
for this source (see \cite{ylh02} for
details). The data are rather uncertain. Thus we plot the
observational points as thick dots. 

If the interpretation of quiescent emission as the thermal
emission from the NS surfaces is correct, then all five NSs
are heated to the neutrino emission regime 
($L_{\rm dh}^\infty \gg L_\gamma^\infty$).
Since $L_{\rm dh}^\infty$ is reliably determined,
for a known $\dot{M}$, and $L_\gamma^\infty$ is measured,
one can immediately estimate the neutrino luminosity of each source
from Eq.\ (\ref{Main}):
$L_\nu^\infty=L_{\rm dh}^\infty-L_\gamma^\infty$. 
In all our cases $L_\nu^\infty$ is
comparable with $L_{\rm dh}^\infty$ (Fig.\ 1).

As seen from Fig.\ 1,
we can treat NSs in 4U 1608--52 and Aql X--1 as low-mass
NSs with very weak neutrino emission from their cores.
NSs in Cen X--4 and SAX J1808.4--3658 seem to require the enhanced 
neutrino emission and are thus more massive.
The status of NS in KS 1731--26 is less certain 
\cite{ylh02} because
of poorly determined $\dot{M}$.
If the real value of $\dot{M}$ is close to the assumed one,
it may also require
some enhanced neutrino emission.
Similar conclusions have been made, 
particularly, in Refs.\ 
\cite{ur01,cgpp01,rbbpz01,rbbpzu02,rbbpz02,bbc02,wgvm02}
with respect to some of these sources or selected groups.

Let us disregard the SAX source for a moment.
We see that the observational point for Cen X--4
lies above (or near) all three limiting heating curves
for massive NSs (with kaon- or pion condensate,
and nucleon Durca cores). Accordingly, we can treat
the NS in Cen X--4 either
as high-mass NS (with kaon-condensed or quark core) or as
medium-mass NS (with pion-condensed, quark,
or Durca-allowed nucleon core). 
If further observations confirm the current status of
SAX J1808.4--3658,
then we will have the only choice
to treat this NS as a high-mass NS with the
nucleon core (and the NS in Cen X--4 as medium-mass NS
with the nucleon core). This would disfavor the hypothesis
on exotic phases of matter
in NS cores. 

Our generic description of neutrino emission (Sect.\ 4) 
is too flexible 
to fix the position of the transition layer in the
NS cores where a slow emission transforms into a fast one.
Adopting a specific physical picture of NS interiors,
with this position determined by microphysics input,
we would be able to construct the sequences
of heating curves for NSs with different $M$, and
attribute certain values of $M$ to any source
(``weigh'' NSs in SXRTs, as proposed by Colpi et al.\ \cite{cgpp01},
just as in the case of cooling isolated NSs 
\cite{kyg02,ygkp02}).

The assumption that the observed X-ray emission
of SXRTs in quiescence emerges from NS interior is
still an attractive hypothesis. However, the theory
of deep crustal heating is solid.
This phenomenon, produced mainly by pycnonuclear
reactions in the inner NS crusts, 
should occur in accreting NSs leading to observational consequences.

Because of the similarity between the
heating and cooling theories, the observations of cooling isolated
NSs and  NSs in SXRTs can be analyzed
together increasing statistics of the sources and
confidence of the results. Recently
the theory of cooling NSs has been confronted
with observations in Refs.\ \cite{yh02,kyg02,ygkp02}.
Some cooling NSs (first of all, RX J0822--43, and
PSR 1055--52) can be interpreted as low-mass
NSs with strong proton superfluidity in their cores. 
Other sources (first of all, Vela, Geminga, and RX J0205+6449)
seem to require enhanced neutrino emission but the
nature of this emission is uncertain, just as for SXRTs
disregarding SAX J808.4--3658.
In this context, the latter source is now
the only one which disfavors exotic phases of
matter in the NS cores.\\

{\bf Acknowledgments.}
This work was supported in part by
RBRF (grants Nos. 02-02-17668 and 00-07-90183).

\end{document}